\documentstyle[12pt,aaspp4,psfig,amstex]{article}

\def\grb{GRB\thinspace{991216}}
\def\ts{\thinspace}

\begin{document}
 
\title{\large \bf The Enigmatic Radio Afterglow of GRB\ts{991216}}
 
\author{
 D. A. Frail\altaffilmark{1}, 
 E.    Berger\altaffilmark{2},
 T.    Galama\altaffilmark{2},
 S. R. Kulkarni\altaffilmark{2},
 G. H. Moriarty-Schieven\altaffilmark{3},
 G. G. Pooley\altaffilmark{4},
 R.    Sari\altaffilmark{5}, 
 D. S. Shepherd\altaffilmark{1},
 G. B. Taylor\altaffilmark{1}, 
 F.    Walter\altaffilmark{2}
}

\altaffiltext{1}{National Radio Astronomy Observatory, P.~O.~Box O,
  Socorro, NM 87801}
  
\altaffiltext{2}{California Institute of Technology, 
Owens Valley Radio Observatory 105-24, Pasadena, CA 91125}

\altaffiltext{3}{Joint Astronomy Centre, 600 A'Ohoku Place, Hilo, HI
  96720}   

\altaffiltext{4}{Mullard Radio Astronomy Observatory, Cavendish
  Laboratory, Madingley Road, Cambridge CB3 0HE}

\altaffiltext{5}{California Institute of Technology,
 Theoretical Astrophysics  103-33, Pasadena, CA 91125}   

\begin{abstract}
  We present wide-band radio observations spanning from 1.4 GHz to 350
  GHz of the afterglow of \grb, taken from 1 to 80 days after the
  burst.  The optical and X-ray afterglow of this burst were fairly
  typical and are explained by a jet fireball. In contrast, the radio
  afterglow is unusual in two respects: (a) the radio light curve does
  not show the usual rise to maximum flux on timescales of weeks and
  instead appears to be declining already on day 1 and (b) the power
  law indices show significant steepening from the radio through the
  X-ray bands.  We show that the standard fireball model, in which the
  afterglow is from a forward shock, is unable to account for (b) and
  we conclude that the bulk of the radio emission must arise from a
  different source. We consider two models, neither of which can be
  ruled out with the existing data. In the first (conventional) model,
  the early radio emission is attributed to emission from the reverse
  shock as in the case of GRB 990123.  We predict that the prompt
  optical emission would have been as bright (or brighter) than 8th
  magnitude. In the second (exotic) model, the radio emission
  originates from the forward shock of an isotropically energetic
  fireball ($10^{54}$ erg) expanding into a tenuous medium (10$^{-4}$
  cm$^{-3}$). The resulting fireball would remain relativistic for
  months and is potentially resolvable with VLBI techniques.  Finally,
  we note that the near-IR bump of the afterglow is similar to that
  seen in GRB 971214 and no fireball model can explain this bump.
\end{abstract}

 
\clearpage

\section {Introduction\label{sec:introduction}}

The intense gamma-ray burst \grb\ was detected on 1999 December 16.67
UT by the Burst and Transient Experiment (BATSE) on board the {\it
  Compton Gamma Ray Observatory} satellite (\cite{kpg99}). Follow-up
observations with the PCA instrument on the {\it Rossi X-ray Timing
  Explorer} (RXTE) satellite resulted in the detection of a previously
uncataloged X-ray source, which was subsequently seen to fade by a
factor of five, seven hours later (\cite{tmm+99}). Uglesich et
al.~(1999\nocite{umh+99}) identified a fading optical source, at a
position consistent with the RXTE transient, and shortly thereafter
the radio counterpart was discovered (\cite{tb99}).

Here we present radio measurements of this burst from 1 GHz to 350
GHz.  While the emission from X-ray and optical afterglow was fairly
typical (\cite{hum+00}), the radio afterglow of \grb\ was unusual in
two respects. First, the onset of the decay began much earlier than
that in most radio afterglows. Second, the temporal decay indices in
the radio, optical and X-ray bands are markedly different from each
other.  We explore a number of possible explanations for these
behaviors.

\section {Observations\label{sec:obs}} 

\noindent
{\it Very Large Array 
  (VLA\footnotemark\footnotetext{The NRAO is a facility of the
    National Science Foundation operated under cooperative agreement
    by Associated Universities, Inc. NRAO operates both the VLA and
    the VLBA.}):} A log of the observations and flux density
measurements are summarized in Table \ref{tab:Table-VLA}.  We used
J0509+1011 (at 8.46 GHz and 4.86 GHz) and J0530+135 (at 1.43 GHz) for
phase calibration. J0542+498 was used for flux calibration at all
frequencies.


\noindent
{\it Very Long Baseline Array (VLBA$^{6}$):} A single 2-hr observation
was carried out at 8.42 GHz and 2-bit samples of a 64 MHz bandwidth
signal in one hand of polarization were recorded. The nearby
($<1.1^\circ$) calibrator J0509+1011, a core-jet source, was observed
every 3 minutes for delay, fringe rate and fringe phase calibration.
The total flux density of the calibrator was found to be 9.5\% less
than was measured by the VLA on the same day.  Given that the jet of
J0509+1011 is likely to have some extended emission that is not
detectable by the VLBA, it is likely that the absolute flux
calibration of the VLBA is well within its nominal value of 5\%.

The radio afterglow was detected at a position of (epoch J2000)
$\alpha=5^h 9^m 31.2983^s$, $\delta=+11^\circ 17^\prime
7.262^{\prime\prime}$, with (conservative 1-$\sigma$ error of  
0.001$^{\prime\prime}$ in each coordinate). The source is unresolved
with a size of less than 0.001$^{\prime\prime}$.

\noindent
{\it Ryle Telescope:} Observations at 15 GHz with the Ryle Telescope
at Cambridge (UK) were made by interleaving 15 minute scans of \grb\
with short scans of the phase calibrator J0509+1011. The flux density
scale was tied to observations of 3C\ts{48} and 3C\ts{286}.

\noindent
{\it Owens Valley Radio Observatory Interferometer (OVRO):} The source
was observed for a single 13 hr track in two continuum 1~GHz bands
(central frequencies 98.481~GHz and 101.481~GHz) under good 3-mm
weather conditions.  Gain calibration used the quasar 0528$+$134, while
observations of Uranus and 3C~454.3 provided the flux density
calibration scale with an estimated uncertainty of $\sim 20$\%.  See
Shepherd et al.~(1998)\nocite{sfkm98}, for details of the calibration
and imaging.  No source was detected (see Table \ref{tab:Table-VLA}).

\noindent
{\it James Clark Maxwell Telescope (JCMT\footnotemark\footnotetext{The
    JCMT is operated by The Joint Astronomy Centre on behalf of the
    Particle Physics and Astronomy Research Council of the UK, the
    Netherlands Organization for Scientific Research, and the National
    Research Council of Canada.}):} Observations in the 350 GHz band
were made using the Sub-millimeter Common-User Bolometer Array
(\cite{hrg+99}).  The data were taken under good sky conditions on
both nights. For flux calibration we used the source CRL618, and
assumed its flux density to be 4.57 $\pm$ 0.21 Jy.  The pointing was
monitored and found to vary by less than 2\arcsec.  See Kulkarni et
al. (1999\nocite{kfs+99}) for details of data reduction.  The source
was not detected at either epoch (see Table \ref{tab:Table-VLA}). At
the position of \grb\ we derive an average flux of $-0.28 \pm 1.1$
mJy.

\section {Results\label{sec:results}}

\begin{figure*}[tb]
  \centerline{\psfig{file=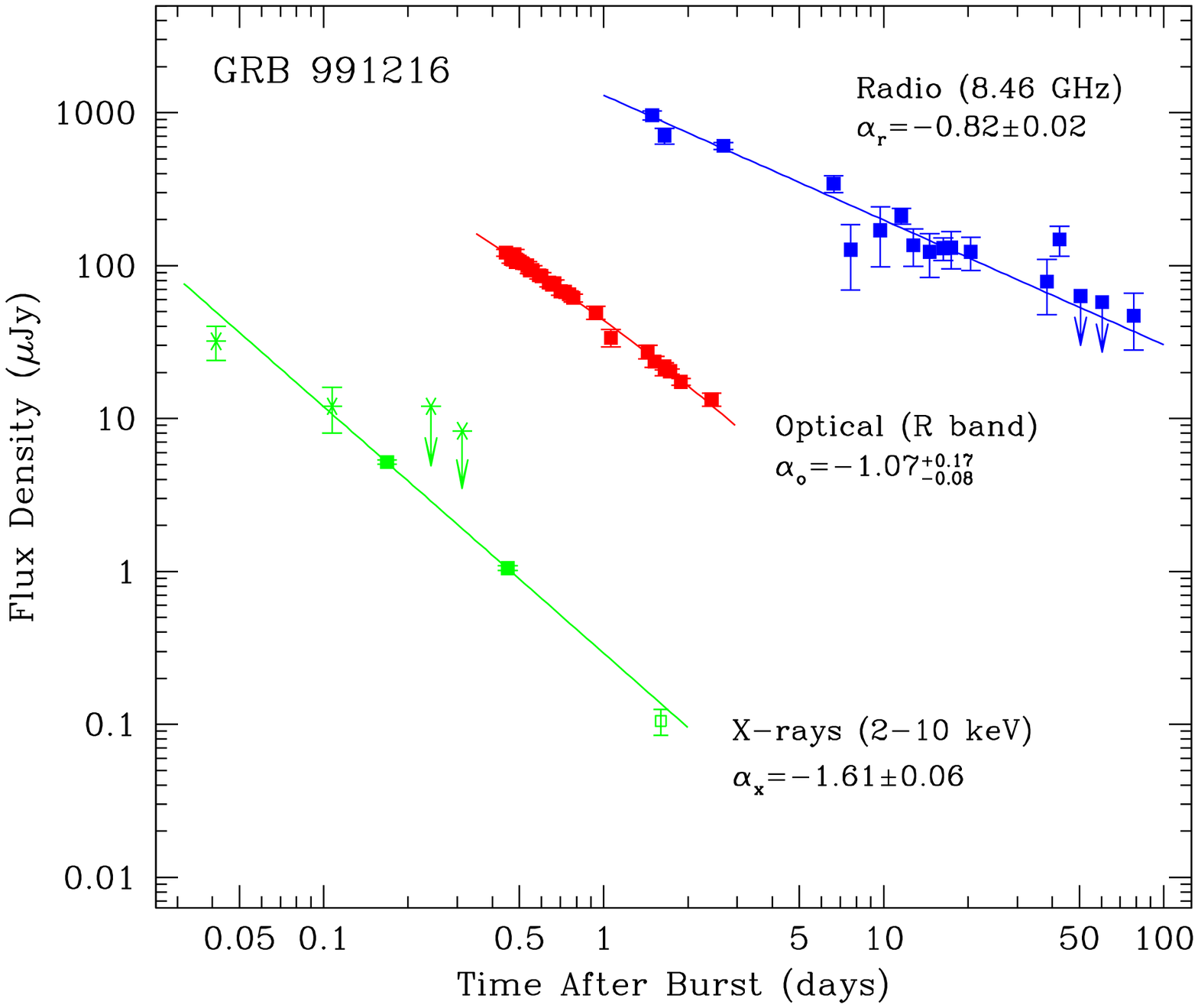,angle=0,width=16cm}}
  \caption[]{Broad-band light curves of \grb. Upper limits
    are plotted as the peak flux density at the location of the
    afterglow plus two times the rms noise in the image. The R-band
    data are taken from Halpern et al. (2000). Optical magnitudes were
    converted to Jansky flux units (\cite{fsi95}) but no correction
    has been made for Galactic extinction. The X-ray data are
    measurements taken by the ASM ($\ast$) and PCA (filled squares)
    instruments on RXTE (\cite{cm99}, \cite{tmm+99}), and the Chandra
    X-ray Observatory ($\square$; \cite{pgg+99}).  X-ray fluxes are
    converted to Janskys using the X-ray slope $\beta_x=-1.1$ derived
    by Takeshima et al.~(1999\nocite{tmm+99}).  The solid lines are
    noise-weighted least squares fits to the data, with the slopes
    $\alpha_\nu$ as indicated (see text for
    details).\label{fig:lc-lcurve}}
\end{figure*}

In Figure \ref{fig:lc-lcurve} we display the 8.46 GHz light curve, as
well as the X-ray and optical (R-band) light curves obtained from
measurements reported in the GRB Coordinates Network
(GCN)\footnotemark\footnotetext{$\rm{http://lheawww.gsfc.nasa.gov/
    docs/gamcosray/legr/bacodine/gcn\_main.html.}$} and Halpern et al.
(2000\nocite{hum+00}). A noise-weighted least squares fit of the form
$F_\nu\propto t^{\alpha_\nu}$ was made to each of these light curves.
Using all the 8.46 GHz data, including the upper limits, we derive
$\alpha_r=-0.82\pm 0.02$ ($\chi^2_r=26.5/15$; here $\chi^2_r$ is the
reduced $\chi^2$).

A similar least squares fit of the optical and X-ray data over the
first three days (Figure \ref{fig:lc-lcurve}), yields
$\alpha_o=-1.33\pm0.01$ ($\chi^2_r=11/28$) and $\alpha_x=-1.61\pm
0.06$ ($\chi^2_r=7.7/3$). From a more extensive dataset, Halpern et
al.~(2000) fit $\alpha_o=-1.07^{+0.17}_{-0.08}$ over the same time
range. We will use their value of $\alpha_o$ in the discussion to
follow.  The relatively large value of $\chi^2_r$ for fit to the X-ray
data presumably reflects the uncertainties inherent in converting the
counts measured by three different instruments (RXTE-ASM, RXTE-PCA and
Chandra ACIS) into Jansky flux units.  Using the RXTE-PCA data alone
avoids this cross-calibration issue and yields $\alpha_x=-1.61\pm
0.05$.

\section{The Unusual Nature of the Radio Afterglow: 
The Failure of the Basic Afterglow Model\label{sec:fail}}

The radio afterglow from \grb\ is unusual on two counts.  First, the
radio afterglow in the centimeter band does not show the usual rise to
a peak value $f_m$ (at epoch $t_m$) before undergoing a power law
decay.  The radio flux appears to decline continuously starting from
the epoch of the first observation. Thus $t_m<1.49$ d as compared to
the 10--100 d seen in other bursts (e.g.  \cite{fwk00}, \cite{fkb+99},
\cite{fks+99}).  Second, the temporal decay indices ($\alpha_\nu$) in
the radio, optical and X-ray bands are markedly different from each
other. Proceeding from radio to higher frequencies, $\alpha_{\nu}$ steepens
by $\sim$0.4 every four decades in frequency.

In contrast, the optical and X-ray afterglow appears to find a
straightforward explanation in the standard afterglow model in which a
jet geometry is invoked (\cite{hum+00}). Below we show that the radio
observations cannot be reconciled with a standard jet (or sphere)
afterglow model. We then explore possible modifications to the
standard model.

The simplest afterglow model is one in which the broad-band afterglow
emission arises from the forward shock of a relativistic blast wave
propagating into a constant density medium (\cite{spn98}).  It is
assumed that the electrons in the forward shock region are accelerated
to a power law distribution for $\gamma_e>\gamma_m$,
$dN/d\gamma_e\propto \gamma_e^{-p}$; here $\gamma_e$ is the Lorentz
factor of the electrons, $p$ is the power law index and $\gamma_m$ is
the minimum Lorentz factor.  Gyration of these electrons in strong
post-shocked magnetic fields gives rise to broad-band afterglow emission.  Two
modifications to this picture are routinely considered. (1) An
inhomogeneous circumburst medium (specifically, $\rho(r)\propto
r^{-2}$; here $\rho$ is the density at distance $r$ from the source).
Such a circumburst medium is expected should GRBs originate from
massive stars (\cite{cl99b}). (2) A jet-like geometry for the blast
wave (\cite{sph99}). This modification is motivated by the propensity
of jets in astrophysical sources.

Regardless of these modifications, the broad band spectrum is composed
of three characteristic frequencies: $\nu_a$, the synchrotron
self-absorption frequency; $\nu_m$, the frequency at which the
emission peaks (and attributed to the electrons with Lorentz factor
$\gamma_m$), and $\nu_c$, the cooling frequency.  Electrons radiating
photons with frequency $>\nu_c$ cool on timescales faster than the age
of the blast wave. The evolution of these frequencies is determined by
the dynamics of the blast wave.  The usual ordering of these
frequencies at epochs relevant to the discussion here is (going from
low to high frequencies) $\nu_a$, $\nu_m$ and $\nu_c$.

For \grb\ the early radio decay implies that $\nu_m$ is already below
the centimeter radio band at 1.49 days.  The steepening of the
afterglow emission from optical to X-ray can be explained by placing
$\nu_c$ between the optical and X-ray bands.  The expected steeping
$\Delta\alpha$ is 1/4 which is marginally consistent with
$\alpha_o-\alpha_x=0.54^{+0.18}_{-0.10}$. However, even if we ignore
this, we are simply unable to explain the decay in the radio band,
since no additional steepening is expected between $\nu_m$ and
$\nu_c$.

The standard afterglow model can be made to agree with the light
curves by postulating an energy slope $p$ which gradually steepens
with increasing electron energy $\gamma_e$.  We use the spherical,
constant density afterglow model (\cite{spn98}) to convert, in each
band, the observed decay index to $p$ and obtain: $p=2.09\pm0.03$
(radio), $p=2.43^{+0.23}_{-0.11}$ (optical), and $p=2.81\pm0.08$ (if
$\nu_c$ is below the X-ray band) or $p=3.15$ (if $\nu_c$ above the
X-ray band).  We are justified in applying the spherical model for
early times ($t<t_J\sim$ 2--5 d) since the jet geometry is manifested
only for $t>t_J$ (\cite{hum+00}).

Curvature is both observed and modeled in the synchrotron spectra of
the non-relativistic shocks from supernova remnants which are
accelerating cosmic rays (e.g. Baring et al.~1999\nocite{ber+99}). To
date, models of ultra-relativistic shocks favor a universal value of
$p$, independent of energy (\cite{vie00}, \cite{gakg00}), but
non-linear effects have yet to be treated.

Nonetheless, the invocation of curvature in the energy distribution of
the electrons cannot explain the observed broad-band spectrum
(Figure~\ref{fig:broadband}) of the afterglow on December 18
(corresponding to 1.33 days after the burst). A plausible fit to the
entire data is obtained with $\nu_a=1.3$~ GHz, $\nu_m=270$~GHz and
$\nu_c=7\times 10^{16}$~Hz and $f_m=3.4$ mJy; this fit is displayed by
the dashed line in Figure~\ref{fig:broadband}.  As the blast wave
slows down, $\nu_m$ moves to lower values while preserving $f_m$ and
thus we expect the flux in the centimeter band to rise, whereas the
observed flux falls. If the afterglow has a jet-like geometry then the
radio afterglow is expected to rise until the epoch $t_J$, and
subsequently decay very slowly ($f_\nu\propto t^{-1/3}$) until $\nu_m$
passes through the centimeter band, after which we expect to see a
decline similar to that seen in the optical ($f_\nu \propto t^{-2.2}$)
(\cite{hbf+99}). As can be seen from Figure~\ref{fig:broadband}, the
radio observations are grossly inconsistent with these expectations,
particularly the decay is much faster than $t^{-1/3}$.

To summarize, while the optical and X-ray observations can be
accounted for by a jet model, the radio observations are inconsistent
with the standard model. This forces us to consider afterglow models
in which the radio emission (at least in bulk) arises from a source
other than the usual forward shock.

\section{A Forward and Reverse Shock Model\label{sec:reverse}}

The most natural explanation for two components would be an early
contribution from a reverse shock followed by a forward shock element
at later times. This is the explanation invoked to account for the
early (1-2 day) radio emission from the afterglow of GRB\ts{990123}
(\cite{sp99b}, \cite{kfs+99} but see \cite{gbw+99}).  The two bursts
share several common features: in both cases, a jet has been deduced
with $t_J\sim$ few days, both were quite bright at gamma-ray energies
and finally both had a seemingly small value of $t_m$ (as measured in
the centimeter band). However, in the case of GRB\ts{990123}, the peak
flux of the forward shock was $f_m<260\,\mu$Jy (\cite{kfs+99}) and the
radio light curve was dominated by the reverse shock. In contrast, the
forward shock for \grb\ appears to be quite strong. This difference
then explains the seemingly different radio light curves.

At late times (i.e. timescales greater than the duration of the burst)
the flux from the reverse shock is expected to fall as $t^{-1.8}$
(\cite{ks00}).  In contrast, the forward shock emission rises as
$t^{1/2}$ for $t<t_J$ and then slowly decays, $\propto t^{-1/3}$ until
the $\nu_m$ moves into the centimeter band. Since $t_J$ is known from
optical observations (\cite{hum+00}), the remaining unknowns are the
strength of the reverse and forward shock emission.

\begin{figure*}[tb]
  \centerline{\psfig{file=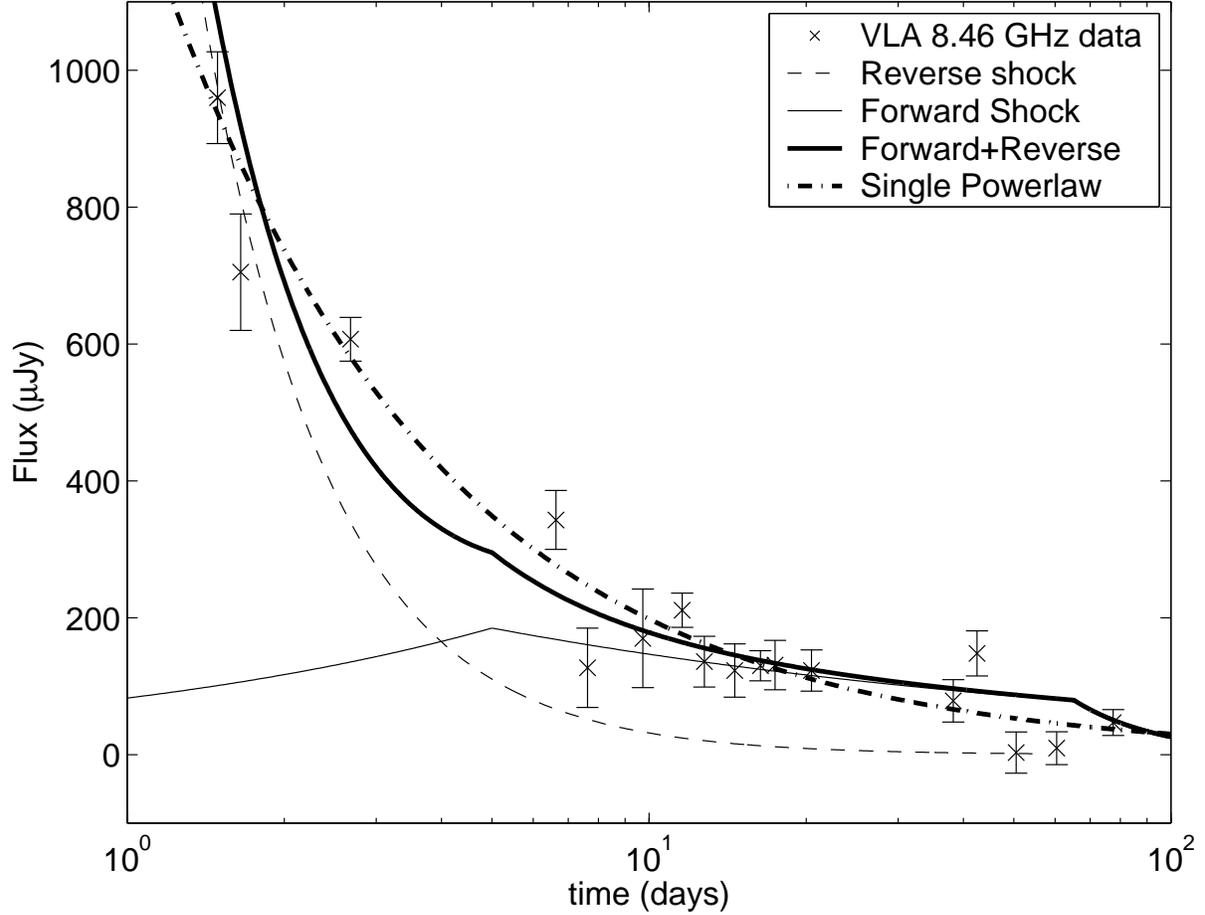,angle=0,width=16cm}}
  \caption[]{Observed and model light curves at 8.46 GHz. The
    dot-dashed line is the power-law fit from Figure
    \ref{fig:lc-lcurve}. The thick solid line is the two component
    model discussed in \S\ref{sec:reverse}, consisting of a reverse
    shock (dashed line) and a forward shock (thin solid line). See
    text for more details.
\label{fig:r-fmodel}}
\end{figure*}

In this picture, the reverse shock dominates the radio emission for
the first few days and the model fit consists of mainly fitting a
power law with $f_\nu\propto t^{-1.8}$. We note that at day 1.5, the
VLA 8.46 GHz flux and the Ryle 15 GHz flux are comparable. This suggests
that the reverse shock is already optically thin at 8.46 GHz at this
epoch -- similar to the situation for GRB 990123 (\cite{kfs+99}).  We
deduce the parameters of the forward shock by fitting the radio to
optical spectrum around $t_J$=5 days to the forward shock model (the
contribution of the reverse shock is expected to be negligible thanks to
the steep decay and since $t$ is comparable to $t_J$, 
the spherical fireball model is still applicable);
we find $\nu_m \sim 1.4\times 10^{12}$ Hz and $f_m=1$ mJy. As can be
seen from Figure~\ref{fig:r-fmodel} this reverse-forward model
provides a reasonable fit to the observations.

There are two predictions of this model. First, we expect $\nu_m$ to
cross the centimeter band at $t_b=t_J(\nu_m/8.46\,{\rm GHz})^{1/2}
\sim 64$ d.  For $t>t_b$, we expect the radio flux to decline as
steeply as the optical flux does for $t>t_J$.  The low flux values as
measured at the VLA around this epoch are in agreement with this
model.

A second prediction is that for $t<t_b$, we expect, the spectrum to
rise as $\nu^{1/3}$ for $\nu <8.46$ GHz.  Unfortunately, the data are
too sparse to rigorously test this expectation. Nonetheless, we note
that at day 17.44, the spectrum between 1.43 and 8.46 GHz can be
described by a simple power-law with slope $\beta_r=-0.45$, steeper
than a $\nu^{1/3}$ slope by 3.6$\sigma$. We consider this to be the
weakest point of the model but do not consider it fatal since the
quoted uncertainties include only instrumental errors and do not
include external effects such as interstellar scintillation.

The strongest confirmation of this model would have been the detection
of an optical flash, as in the case of GRB 990123 (\cite{abb+99}).
The strong radio emission from the reverse shock allows us to predict
(by scaling from the optical and radio observations of GRB 990123,
\cite{sp99b}) that the flash would have been 8th magnitude or even
brighter. Unfortunately, this event occurred during daytime and
therefore was not observed by existing prompt optical counterpart
experiments (LOTIS -- H. S. Park; ROSTE -- C. Akerlof; {\it pers.
  comm.})

We end this section by noting a worrisome and puzzling issue: we are
unable to provide a consistent explanation for the near-IR, optical
and X-ray observations with a standard fireball afterglow spectrum.
As noted in Figure~\ref{fig:broadband}, there is a broad maximum
around $2\times 10^{14}$ Hz -- suggesting that this is the peak
frequency ($\nu_m$) of the fireball. Fitting a template afterglow
spectrum we obtain the following: $\nu_{m1}=2.1\pm 0.6 \times 10^{14}$
Hz, $f_{m1}=150\pm 10$ $\mu$Jy and $\nu_{c1}=2\times 10^{16}$ Hz.  We
note that a similar broad peak in the near IR (and attributed to
$\nu_m \sim 3\times 10^{14}$ Hz at $\Delta t=0.5$ d) was observed for
GRB 971214 (\cite{rkf+98}). However, if we evolve this $\nu_m$ back in
time (with $\nu_m\propto t^{-3.2}$) we predict a {\it rising} R-band
light curve, inconsistent with the observations (Figure
\ref{fig:lc-lcurve}). Moving $\nu_m$ to lower frequencies solves this
problem but we are left with no explanation for the ``near-IR'' bump.


\section{A Two-Component Forward Shock Model\label{sec:forward}}

We now consider a model in which much of the radio emission arises as
the forward shock of an additional fireball (hereafter the second
fireball). The principal attraction of the second fireball is that we
no longer need to relate the radio decay rate to those at optical and
radio frequencies. We clarify that the optical and X-ray observations
are explained by the forward shock of the fireball (the first
fireball) discussed in the previous section.  As noted earlier, there
is good evidence suggesting that the first fireball is a jet. Thus the
second fireball must be a more isotropic fireball and move at a
smaller Lorentz factor. Indeed, in some GRB models, the central engine
is expected to inject two fireballs: a high $\Gamma$ jet and a low
$\Gamma$ spherical wind.

\begin{figure*}[tb]
\centerline{\psfig{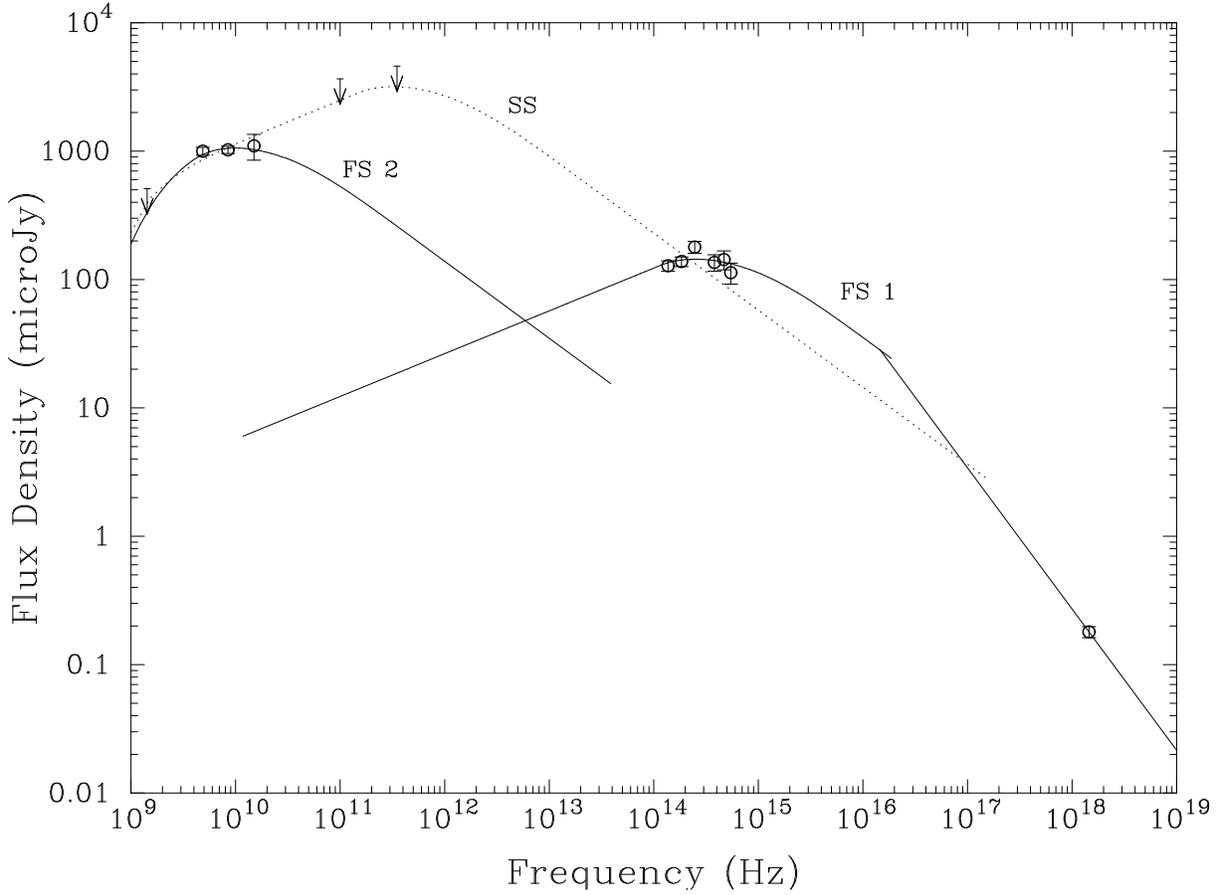}}
\vskip0.0truein
\caption[]{Radio to X-ray spectral flux distribution of \grb\ on 1999 December
  18.00 ($\Delta t$=1.33 days after the burst). Optical and infrared
  measurements are taken from Halpern et al. (2000) and are
  interpolated to this epoch assuming the decay rate of
  $\alpha_o=-1.07^{+0.17}_{-0.08}$, as measured by Halpern et al. at
  early times. The optical/IR data have been corrected for Galactic
  foreground extinction (\cite{sfd98}), giving E(B$-$V)=0.634 with an
  an uncertainty of 10\%. The 1.4 GHz upper limit (plotted as three
  times the rms noise) and the 4.8 GHz data point were taken at the
  Westerbork Synthesis Radio Telescope (WSRT) by Rol et
  al.~(1999)\nocite{rvs+99}.  The flux density at 8.46 GHz derived by
  extrapolating of the power-law decay in Figure~\ref{fig:lc-lcurve}.
  The upper limits at 100 GHz and 350 GHz have been extrapolated back
  to this epoch by assuming a worst case decay rate of
  $\alpha_o-2\sigma$.  The dotted and solid lines are fits to the data
  for a synchrotron spectrum from a relativistic blast wave as
  specified by Granot et al.~(1999a\nocite{gps99b}; see their Figure
  10 for the equipartion field model). We assume $p=2.2$ and scale it
  by $\nu_m$ and $f_m$ to derive a function $g(\nu)$ with asymptotic
  limits of $\nu^{1/3}$ and $\nu^{(p-1)/2}$. We account for
  synchrotron self-absorption at $\nu_a$ by multiplying $g(\nu)$ by
  F$_\nu=[1-exp(-\tau)]/\tau$, where $\tau=(\nu/\nu_{a})^{-5/3}$
  (\cite{gps99c}).
\label{fig:broadband}}
\end{figure*}

A reasonable fit to the radio data of this second fireball (FS 2; see
Figure~\ref{fig:broadband}) on day 1.33 is provided by
$f_{m2}\simeq{1.2}$ mJy, $\nu_{m2}$=7 GHz, and $\nu_{a2}$=2 GHz.  The
location of the cooling frequency $\nu_{c2}$ is unconstrained. As a
test, we evolved the afterglow spectrum forward in time. The model
does an excellent job reproducing the declining flux density from 1.43
and 8.46 GHz at 17.44 days (an observation which the reverse-forward
shock model fails to explain), but at day 60.40 it predicts a 1.43 GHz
flux of $\sim{100}$ $\mu$Jy, where only an upper limit of $-57\pm44$
$\mu$Jy is measured.  Again we consider this 3-$\sigma$ discrepancy as
a major, but not fatal, weakness of this model.

The three inferred parameters ($\nu_{m2}$, $f_{m2}$, $\nu_{a2}$) allow
us to obtain the energy of the blast wave and the density of the
ambient medium (\cite{wg99}): $E_{52}\sim 10^2$ erg and $n\sim
10^{-4}$ cm$^{-3}$; these values are relatively insensitive to the
value of the unknown $\nu_c$ (which is however constrained to lie
above the optical band). The large $E$ and small $n$ are primarily due
to the small value of $t_m$.

If this interpretation is correct then we have uncovered the first
example of a GRB exploding in a very low density medium -- perhaps the
halo of a host galaxy.  The dynamics of explosions is governed by the
ratio $E/n$, and as noted above, this ratio is perhaps $10^5$ larger
than that typically derived in other afterglow.  For this reason, both
fireballs, the high $\Gamma$ and the low $\Gamma$ fireballs, would
then be expanding at high Lorentz factors days after the burst.  The
$\Gamma$ for the low $\Gamma$ fireball would be an impressive 20 one
day after the burst, and the fireball would have had a size of 100
microarcseconds three weeks after the burst -- almost within reach of
measurable with VLBI techniques (cf. \cite{tfbk97}). The jet fireball
would be expanding even faster in which case the opening angle of the
jet is not 6$^\circ$ (\cite{hum+00}) but only 1$^\circ$.

To summarize, the radio afterglow of \grb\ is unusual and cannot be
explained by the standard forward shock model. A conventional
reverse-forward shock model or an exotic two-component forward shock
model can account for the observations, but each has one major (but not
necessarily fatal) weakness.  Finally, we have no explanation for the
near-IR bump seen on day 1.33. \grb\ shows that there may be yet new
surprises in GRB afterglows.

\acknowledgements DAF thanks Chris Fassnacht, Steve Myers, Lin Yan
and Jim Ulvestad for generously giving up portions of their VLA time
so that GRB\ts{991216} could be observed. DAF thanks J. Halpern for
making his paper available prior to publication and Y. Gallant and M.
Vietri for useful discussions.  We would like to thank S.~Jogee for
preparing the first observations of this burst at the Owens Valley
Observatory and A. Sargent for generously allocating the time on short
notice.  Research at the Owens Valley Radio Observatory is supported by
the National Science Foundation through NSF grant number AST 96-13717.
SRK's research is supported by grants from NSF and NASA. RS and TJG are
supported by Sherman Fairchild Fellowships.

\clearpage
  

\begin{thebibliography}{}

\bibitem[{Akerlof} {\it et al.}  1999]{abb+99}
{Akerlof}, C. {\it et al.}  1999, Nature, 398, 400.

\bibitem[{Baring} {\it et al.}  1999]{ber+99}
{Baring}, M.~G., {Ellison}, D.~C., {Reynolds}, S.~P., {Grenier}, I.~A., and
  {Goret}, P. 1999, ApJ, 513, 311.

\bibitem[{Chevalier} \& {Li} 1999]{cl99b}
{Chevalier}, R.~A. and {Li}, Z.-Y. 1999, ApJ (Let) submitted;astro-ph/9908272.

\bibitem[Corbet \& Smith 1999]{cm99}
Corbet, R. and Smith, D. 1999, GCN notice 506.

\bibitem[{Frail} {\it et al.}  1999]{fks+99}
{Frail}, D. {\it et al.}  1999, ApJ.
\newblock ApJ submitted; astro-ph/9910060.

\bibitem[Frail {\it et al.}  1999]{fkb+99}
Frail, D.~A. {\it et al.}  1999, ApJ (Let) in press; astro-ph/9909407.

\bibitem[Frail, Waxman \& Kulkarni 2000]{fwk00}
Frail, D.~A., Waxman, E., and Kulkarni, S.~R. 2000, ApJ in press;
  astro-ph/9910319.

\bibitem[{Fukugita}, {Shimasaku} \& {Ichikawa} 1995]{fsi95}
{Fukugita}, M., {Shimasaku}, K., and {Ichikawa}, T. 1995, PASP, 107, 945.

\bibitem[{Galama} {\it et al.}  1999]{gbw+99}
{Galama}, T.~J. {\it et al.}  1999, Nature, 398, 394.

\bibitem[Gallant {\it et al.}  2000]{gakg00}
Gallant, Y.~A., Achterberg, A., Kirk, J.~G., and Guthmann, A.~W. 2000, 5th
  Huntsville Gamma-Ray Burst Symposium, in press, astro-ph/0001509.

\bibitem[{Granot}, {Piran} \& {Sari} 1999a]{gps99b}
{Granot}, J., {Piran}, T., and {Sari}, R. 1999a, ApJ, 513, 679.

\bibitem[{Granot}, {Piran} \& {Sari} 1999b]{gps99c}
{Granot}, J., {Piran}, T., and {Sari}, R. 1999b, ApJ in press;
  astro-ph/9908007.

\bibitem[Halpern {\it et al.}  2000]{hum+00}
Halpern, J.~P., Uglesich, R., Mirabal, N., Kassin, S., {\it et al.}  2000, ApJ
  (Let) submitted.

\bibitem[{Harrison} {\it et al.}  1999]{hbf+99}
{Harrison}, F.~A. {\it et al.}  1999, ApJ, 523, L121.

\bibitem[{Holland} {\it et al.}  1999]{hrg+99}
{Holland}, W.~S. {\it et al.}  1999, mnras, 303, 659.

\bibitem[Kippen, Preece \& Giblin 1999]{kpg99}
Kippen, R.~M., Preece, R.~D., and Giblin, T. 1999, GCN notice 463.

\bibitem[{Kobayashi} \& {Sari} 2000]{ks00}
{Kobayashi}, S. and {Sari}, R. 2000, astro-ph/9910241.

\bibitem[{Kulkarni} {\it et al.}  1999]{kfs+99}
{Kulkarni}, S.~R. {\it et al.}  1999, ApJ, 522, L97.

\bibitem[Piro {\it et al.}  1999]{pgg+99}
Piro, L., Garmire, G., Garcia, M., Marshall, F., and Takeshima, T. 1999, GCN
  notice 500.

\bibitem[Ramaprakash {\it et al.}  1998]{rkf+98}
Ramaprakash, A.~N. {\it et al.}  1998, Nature, 393, 43.

\bibitem[Rol {\it et al.}  1999]{rvs+99}
Rol, E., Vreeswijk, P.~M., Strom, R., Kouveliotou, C., Pian, E., Castro-Tirado,
  A., Hjorth, J., and Greiner, J. 1999, GCN notice 491.

\bibitem[{Sari} \& {Piran} 1999]{sp99b}
{Sari}, R. and {Piran}, T. 1999, ApJ, 517, L109.

\bibitem[{Sari}, {Piran} \& {Halpern} 1999]{sph99}
{Sari}, R., {Piran}, T., and {Halpern}, J.~P. 1999, ApJ, 519, L17.

\bibitem[{Sari}, {Piran} \& {Narayan} 1998]{spn98}
{Sari}, R., {Piran}, T., and {Narayan}, R. 1998, ApJ, 497, L17.

\bibitem[Schlegel, Finkbeiner \& Davis 1998]{sfd98}
Schlegel, D.~J., Finkbeiner, D.~P., and Davis, M. 1998, ApJ, 500, 525.

\bibitem[Shepherd {\it et al.}  1998]{sfkm98}
Shepherd, D.~S., Frail, D.~A., Kulkarni, S.~R., and Metzger, M.~R. 1998, ApJ,
  497, 859.

\bibitem[Takeshima {\it et al.}  1999]{tmm+99}
Takeshima, T., Markwardt, C., Marshall, F., Giblin, T., and Kippen, R.~M. 1999,
  GCN notice 478.

\bibitem[Taylor \& Berger 1999]{tb99}
Taylor, G.~B. and Berger, E. 1999, GCN notice 483.

\bibitem[Taylor {\it et al.}  1997]{tfbk97}
Taylor, G.~B., Frail, D.~A., Beasley, A.~J., and Kulkarni, S.~R. 1997, Nature,
  389, 263.

\bibitem[Uglesich {\it et al.}  1999]{umh+99}
Uglesich, R., Mirabal, N., Halpern, J., Kassin, S., and Novati, S. 1999, GCN
  notice 472.

\bibitem[Vietri 2000]{vie00}
Vietri, M. 2000, ApJ (Let) submitted, astro-ph/0002269.

\bibitem[{Wijers} \& {Galama} 1999]{wg99}
{Wijers}, R. A. M.~J. and {Galama}, T.~J. 1999, ApJ, 523, 177.

\end{thebibliography}

\clearpage
 
\newpage 
\begin{deluxetable}{lrccc}
\tabcolsep0in\footnotesize
\tablewidth{\hsize}
\tablecaption{Radio Observations of \grb\label{tab:Table-VLA}}
\tablehead {
\colhead {Epoch}      &
\colhead {$\Delta t$} &
\colhead {Telescope}      &
\colhead {$\nu_\circ$}      &
\colhead {S $\pm$ $\sigma$} \\
\colhead {(UT)}      &
\colhead {(days)}    &
\colhead { }    &
\colhead {(GHz)}    &
\colhead {($\mu$Jy)}
}
\startdata
1999 Dec. 18.00  & 1.33  & Ryle & 15.0 & 1100$\pm$250 \\
1999 Dec. 18.16  & 1.49  & VLA  & 8.46 & 960$\pm$67   \\
1999 Dec. 18.32  & 1.65  & VLBA & 8.42  & 705$\pm$85   \\
1999 Dec. 18.48  & 1.81  & JCMT & 350  & 650$\pm$1560 \\
1999 Dec. 19.30  & 2.63  & OVRO & 99.9 & 90$\pm$700  \\
1999 Dec. 19.35  & 2.68  & VLA  & 8.46 & 607$\pm$32   \\
1999 Dec. 19.45  & 2.78  & JCMT & 350  & $-2000\pm$1670 \\
1999 Dec. 20.09  & 3.42  & Ryle & 15.0 & $-100\pm$400 \\
1999 Dec. 22.01  & 5.34  & Ryle & 15.0 & $-10\pm$200 \\
1999 Dec. 23.30  & 6.63  & VLA  & 8.46 & 343$\pm$43   \\
1999 Dec. 24.29  & 7.62  & VLA  & 8.46 & 127$\pm$58   \\
1999 Dec. 26.40  & 9.73  & VLA  & 8.46 & 170$\pm$72   \\
1999 Dec. 28.24  & 11.57 & VLA  & 8.46 & 211$\pm$25   \\
1999 Dec. 29.43  & 12.76 & VLA  & 8.46 & 136$\pm$37   \\
1999 Dec. 31.26  & 14.59 & VLA  & 8.46 & 123$\pm$39   \\
2000 Jan. \phantom{0}2.01  & 16.34 & VLA  & 8.46 & 130$\pm$22   \\
2000 Jan. \phantom{0}3.11  & 17.44 & VLA  & 8.46 & 131$\pm$36   \\
2000 Jan. \phantom{0}3.11  & 17.44 & VLA  & 4.86 & 126$\pm$31   \\
2000 Jan. \phantom{0}3.11  & 17.44 & VLA  & 1.43 & 257$\pm$51   \\
2000 Jan. \phantom{0}6.15  & 20.48 & VLA  & 8.46 & 123$\pm$30   \\        
2000 Jan. 23.95 & 38.28 & VLA  & 8.46 & 79$\pm$31   \\        
2000 Jan. 28.16 & 42.49 & VLA  & 8.46 & 148$\pm$33    \\
2000 Feb. 05.18 & 50.51 & VLA  & 8.46 & 3.1$\pm$30    \\
2000 Feb. 15.07 & 60.40 & VLA  & 1.43 & $-$57$\pm$44  \\
2000 Feb. 15.07 & 60.40 & VLA  & 8.46 & 9.6$\pm$24    \\
2000 Mar. 03.85 & 78.18 & VLA  & 8.46 & 47.0$\pm$19   \\
\enddata 
\tablecomments{The columns are (left to right), (1) UT date of the
  start of each observation, (2) time elapsed since the GRB 991216
  event (i.e. $t_\circ$=1999 December 16.67 UT), (3) telescope name,
  (4) observing frequency, (5) flux density of the radio transient,
  with the error given as the rms noise on the image. The epoch on
  Jan. 23.95 UT is an average of two days of data (Jan. 21.95 UT and
  Jan. 25.94 UT). All VLA observations were made when it was in
  its ``B-array'' configuration.}
\end{deluxetable}

\clearpage
 
\clearpage

\clearpage

\end{document}